\documentclass[aps,prl,reprint,twocolumn]{revtex4}

\usepackage{graphicx}
\usepackage{amsmath}
\usepackage{braket}
\usepackage{xcolor}
\newcommand{\varm}{\mathrm{var}}
\usepackage{hyperref}

\begin{document}

\title{Generalized squeezing as a witness}
 \author{\v{S}imon Br\"auer}
 \affiliation{Department of Optics, Faculty of Science, Palack\'{y} University, 17. Listopadu 12, 77146 Olomouc, Czech Republic}

 \author{Tom\'a\v{s} Opatrn\'y}
 \affiliation{Department of Optics, Faculty of Science, Palack\'{y} University, 17. Listopadu 12, 77146 Olomouc, Czech Republic}

 \author{ Petr Marek}
 \affiliation{Department of Optics, Faculty of Science, Palack\'{y} University, 17. Listopadu 12, 77146 Olomouc, Czech Republic}

\date{\today }

\begin{abstract}
Quantum systems can be prepared in an infinite continuum of states, but only some of them can be used as resources for quantum technologies. Discerning whether a specific quantum state falls into this class, is often a challenging task. We show that it can be performed by looking at the squeezing of the quantum states - a scenario in which the variance of some observable operator is suppressed below the threshold given by the non-useful states. We discuss the general concept first; then we illustrate the approach by evaluating cubic nonlinear squeezing in the system of collective atomic spins.
\end{abstract}

\maketitle

\section{Introduction}
The applications of quantum physics are built upon specific quantum states and their properties. Quantum computing relies on quantum states that have both the computational capacity and potential for fault tolerance \cite{Gottesman2001Jun,baragiola2019,terhal2020,grimsmo2020,bourassa2021,grimsmo2021, Konno2024Jan,Marek2024May}, and quantum metrology employs probes that have heightened sensitivity to some specific form of disturbance \cite{giovannetti2004, giovannetti2006, schnabel2010,oh2020,cao2024, demille2024, len2022, lee2024, brawley2016, gessner2019,jia2024}. The specific property required for the protocol varies; in quantum computation, and more generally for the field of quantum information processing, the desired quantum states arise from the assumptions of the envisioned theoretical model - squeezed states \cite{filip2005,miwa2014} and their superpositions \cite{Gottesman2001Jun,serikawa2018, Konno2024Jan,huang2015}, specific superpositions of Fock states \cite{michael2016, grimsmo2020}, or quantum states produced by some currently unfeasible experimental processes \cite{Marek2011Nov,Marek2018Feb, Miyata2016Feb,sakaguchi2023}. On the other hand, in metrology applications, the desired property is often the variance of the state with respect to some observable \cite{schnabel2010,Aasi2013Aug, VirgoCollaboration2019Dec,gessner2019} - the best established example being the use of quadrature squeezed light for detection of gravitation waves \cite{Aasi2013Aug, VirgoCollaboration2019Dec}, or Fock states \cite{wolf2019,oh2020,deng2024} and their superpositions \cite{dowling2008,mccormick2019} for measuring phase and minute displacements.

Interestingly, these two approaches are not disjunctive as the quantum states required for quantum information processing can be often defined as eigenstates of some observables \cite{Gottesman2001Jun,filip2005, Marek2024May, Brauer2021Jul, Kala2022Aug}. The variance of these operators can often directly determine the amount of noise that is introduced in measurement induced implementations of the protocols \cite{filip2005,Miyata2016Feb,Marek2018Feb,kala2024}. Consequently, the variance of these operators can be used to characterize the quality of approximative realizations of the states. The reduced variance in the specific observable, its squeezing, can be also seen as a quantum resource \cite{Albarelli2018Nov,Chitambar2019Apr} - a feature of quantum states necessary for the applications, which cannot be always feasibly generated, the most prominent examples being quantum entanglement \cite{Sorensen2001May} and quantum non-Gaussianity \cite{Walschaers2021Sep}. Interestingly, the variance reduced below some threshold can be taken as a witness of those resources being present in the state.

This is nicely demonstrated on the case of cubic nonlinearity. For bosonic modes of light, this single deterministic non-Gaussian nonlinear operation is sufficient for optical quantum computation \cite{Lloyd1999Feb, Gottesman2001Jun, sakaguchi2023} together with Gaussian operation. In collective spins, the cubic operation can serve as a resource for quantum metrological protocols \cite{Strobel2014Jul, Pezze2018Sep, Evrard2019May,gessner2019} and counter-diabatic driving for fast preparation of Dicke states \cite{Opatrny2016Feb}. The cubic operation can be deterministically implemented with the help of Gaussian operations and a single non-Gaussian ancillary state \cite{Park2018May, Marek2011Nov, Gottesman2001Jun, Ghose2007Apr}. The ideal resource state is an eigenstate of the cubic operator - nonphysical state that can be prepared only as an approximation \cite{yukawa2013,Yanagimoto2020Jun,Konno2021Feb}. For these approximations, the variance of the cubic operator depends on many things, one of them being the stellar rank of the approximation \cite{chabaud2020,fiurasek2022}. This, in turn means, that the variance of the nonlinear operator can be taken as the witness of some stellar rank or general non-Gaussian properties.

In this paper we extend this concept into a general framework for quantum resources. We introduce a normalized nonlinear squeezing parameter that can be used as a witness to  guarantee presence of some quantum resource. We then discuss the ramifications of this framework in the area of preparing and verifying quantum states, and we demonstrate the concept of cubic squeezing for systems of collective spins.

\section{Squeezing}

In quantum physics, squeezing refers to the process of producing quantum states in which variance with respect to specifically chosen observable is reduced at the cost of increasing variance of the non-commuting observables. In practice, the variance of the tested (squeezed) state $\hat{\rho}_T$ is often compared to some unsqueezed \textit{benchmark state} $\hat{\rho}_B$, often called a \textit{free state}, and the \textit{specific}  squeezing is given by a relative quantity
            \begin{equation}\label{eq:basic variance}
                \xi = \frac{\text{var}_{\hat{\rho}_T}[\hat{\mathcal{O}}]}{\text{var}_{\hat{\rho}_B}[\hat{\mathcal{O}}]},
            \end{equation}
            where $\text{var}_{\hat{\rho}}[\hat{\mathcal{O}}] = \text{Tr}[\hat{\rho} \hat{\mathcal{O}}^2] - (\text{Tr}[\hat{\rho} \hat{\mathcal{O}}])^2$. States in which $\xi = 0$ are eigenstates of operator $\hat{\mathcal{O}}$. Consequently, the ratio (\ref{eq:basic variance}) is a way to quantify the difference between the tested state and the target eigenstate, and it is often directly related to the amount of fluctuations arising in metrological or quantum information processing applications.

            Squeezing can be observed in any quantum system, but it plays the most important role in systems with high Hilbert space dimension, in which it is often not straightforward to prepare eigenstates of operators.
                        
            For the harmonic oscillator with quadrature operators $\hat{x}$ and $\hat{p}$ commuting to $\left[ \hat{x}, \hat{p} \right] = i$, squeezing of any given state $\hat{\rho}_T$ in quadrature operator $\hat{x}$ is defined by equation \eqref{eq:basic variance} where $\hat{\mathcal{O}} = \hat{x}$. The maximally squeezed state is a non-physical infinite energy quadrature eigenstate and the benchmark state is the classical vacuum ground state, $ \hat{\rho}_B = |0\rangle\langle 0|$ with $\varm_{|0\rangle\langle 0|} \hat{x}=1/2$. The normalization term ensures that values $\xi<1$ demonstrate the non-classicality of the state. As a metric, squeezing in quantum harmonic oscillators is much more practical and therefore more widely used than, for example, fidelity or distance to target state \cite{Jozsa1994dec, Braunstein1994May}, which often cannot be effectively evaluated because the target state is not normalized.

            A similar situation exists in the system of collective spins of $N$ atoms \cite{Wineland1992Dec, Kitagawa1993Jun, Ma2011Dec}, which is discrete but approaches continuous nature for large $N$. In such a system, the angular momentum operators $\hat{J}_x$, $\hat{J}_y$, and $\hat{J}_z$ can be used both for theoretical description and practical experimental measurements \cite{Leroux2010Feb, Hines2023Aug, Zhou2020Jul, Marciniak2022Mar, Montenegro2025Jan}. In these scenarios, the spin of the system is aligned into a single direction, meaning, for example, $\langle \hat{J}_x\rangle = \langle \hat{J}_y\rangle = 0$, and the metrological properties depend on second moments of these operators, $\varm\left(\hat{J}_x\right)$, for example. The natural choice of the normalization benchmark is the ground state of the operator $\hat{J}_z$, an analogy of the vacuum state in a harmonic oscillator, for which $\varm \left(\hat{J}_x\right) = \frac{N}{4}$.
           
            Squeezed states are used as resources in quantum protocols. Even more, operations manipulating the system can be divided into those that can generate squeezing and those that can only transform the state without improving the squeezing \cite{Braunstein2005May}. Unitary operators $\hat{U}_F$ that perform such transformations are called \textit{free operations} in quantum resource theories \cite{Yadin2018Dec, Albarelli2018Nov, Chitambar2019Apr, Carrara2020Dec}, and allow us to define \textit{resource} squeezing as
                \begin{equation}\label{eq:resource variance}
                    \xi_{\hat{\mathcal{O}}}(\hat{\rho_T}) = \frac{\mathrm{min}_{\hat{U}_F}\varm_{\hat{\rho_T}} \left(\hat{U}_F^{\dagger}\hat{\mathcal{O}}\hat{U}_F\right)}{\varm_{\hat{\rho_B}} \left(\hat{\mathcal{O}}\right)}.
                \end{equation}
             The free operations always depend on the physical system. For the quadrature squeezing, the sets of free operations are composed of displacement and phase shift in harmonic oscillator and rotation in systems of collective spins, while the benchmark states are coherent states. For squeezing in general observables that can be expressed as nonlinear combination of quadrature or angular momentum operators, the sets of free operations and benchmark states have the largest practical impact when driven by specific experimental limitations. In particular, selecting Gaussian states and Gaussian operations as benchmark states and free operations allows the quantities \eqref{eq:basic variance} and \eqref{eq:resource variance} to witness non-Gaussianity present in the tested states. Note that selecting a class of states, such as Gaussian states, as benchmark states necessitates performing some kind of minimization, but the derived thresholds are then immutable and can be used as straightforward numeric benchmarks.

             In contrast to \eqref{eq:basic variance}, which requires prior knowledge of the specific operator, \eqref{eq:resource variance} quantifies the amount of squeezing in any observable that can be revealed by the free non-squeezing operations if one has full knowledge of the quantum state. The quantity \eqref{eq:resource variance} also cannot increase under such operations and is, therefore, better suited for the description of squeezing as a quantum resource.

             The \textit{specific} squeezing \eqref{eq:basic variance} is practical for optimizing the state preparation protocols, in which the goal is to obtain quantum states best approximating the desired eigenstate of $\hat{\mathcal{O}}$ to be used as a resource \cite{Marek2024May, Carrara2020Dec}. In this case, the preparation procedure is optimized to minimize over the set of preparable states to obtain $\tilde{\xi}_{\mathrm{min}} = \mathrm{min}_{\hat{\rho}}  \tilde{\xi}_{\mathcal{O}}(\hat{\rho})$. The \textit{resource} squeezing \eqref{eq:resource variance} can be used as a general quantifier for the chosen property. For example, it can be applied to an experimentally generated state with the goal of determining whether it possesses the required property, either directly or in an extractable form. In some scenarios it is also conceivable to aim for maximal resource squeezing by scrutinizing some set of preparable quantum states to obtain $\xi_{\mathrm{min}} = \mathrm{min}_{\hat{\rho}}  \tilde{\xi}_{\mathcal{O}}(\hat{\rho})$, even though this approach is computationally the most demanding. In all cases, value $\xi = 0$ confirms the investigated state as the perfect eigenstate of the target operator, best suited for any application. On the other hand, $\xi \ge 1$ implies the state is no better than a free state and is therefore unsuitable for further applications. For any quantum state, the intermediate value $0<\xi < 1$ can then serve as a quantifier for the given resource, which can be expressed either directly or in the dB scale, $\xi[\text{dB}] = 10 \log_{10} \xi$.

        \section{Example: cubic squeezing in collective spin systems}
        It was mentioned in the previous section that the nature of generalized squeezing is strongly informed by the specific choice of not only the target operator but also the sets of free states and free operations. Let us demonstrate this in a particular example: the cubic squeezing in a system of collective spins. The cubic operation \cite{Gottesman2001Jun} and the related cubic squeezing \cite{Brauer2021Jul, Yanagimoto2020Jun, Kala2022Aug} was originally investigated for the quantum harmonic oscillator but can also be expanded to the collective spin systems \cite{Opatrny2016Feb, Colombo2022Aug}.
        \sloppy
        The collective spin systems can be described with help of collective spin operators, $\hat{J}_x$, $\hat{J}_y$, and $\hat{J}_z$, with commutator $\left[\hat{J}_k, \hat{J}_l \right] = i \epsilon_{klm}\hat{J}_m$ (where $k,l,m = \left\{ x, y,z\right\}$. We can use Schwinger's representation equivalent to two bosonic modes with annihilation(creation) operators $\hat{a}_{1,2}$($\hat{a}^{\dagger}_{1,2}$) satisfying commutator $[\hat{a}_k,\hat{a}^{\dagger}_l] = \delta_{kl}$. The collective spin operators have the form $\hat{J}_x = \frac{1}{2}\left(\hat{a}^{\dagger}_1\hat{a}_2 + \hat{a}_1 \hat{a}^{\dagger}_2 \right)$, $\hat{J}_y = \frac{1}{2i}\left(\hat{a}^{\dagger}_1\hat{a}_2 - \hat{a}_1 \hat{a}^{\dagger}_2 \right)$ and $\hat{J}_z = \frac{1}{2}\left(\hat{a}^{\dagger}_1\hat{a}_1 - \hat{a}^{\dagger}_2 \hat{a}_2 \right)$.  Since $\hat{J}_{x,y,z}$ commutate with $\hat{N} \equiv \hat{a}_1^{\dagger}\hat{a}_1 + \hat{a}_2^{\dagger}\hat{a}_2$, one can consider a finite-dimensional Hilbert space with a fixed number of particles $N$. In case of a large number of atoms in the system ($N\gg 1$), it is possible to select an area near the pole that appears to be a plane, and thus it is possible to move from a three-dimensional space to a two-dimensional space (also known as Holstein-Primakoff transformation \cite{Holstein1940Dec}). The operator $\hat{J}_z$ thus changes to its mean value $\hat{J}_z \approx \frac{N}{2}$.

        The cubic operation can be implemented by a unitary operator
                \begin{align}\label{eq:cubic_unitary}
                    \hat{U}_c (\chi) = \exp{\left(i\frac{\chi}{3} \left(\frac{N}{2}\right)^{-\frac{3}{2}} \hat{J}_z^3\right)},
                \end{align}
         where the factor $\left(\frac{N}{2}\right)^{-\frac{3}{2}}$ represents the scaling of the effective strength arising from the relation between the commutation relations and dimension of the system. Having this scaling in this definition allows the effective interaction strength $\chi$ to be independent on the system size, which allows for better comparison. The nonlinear cubic squeezing is then tied to variance of operator
         \begin{align}\label{eq:operator_O}
                    \hat{\mathcal{O}}_c(\chi) = \hat{U}_c(\chi) \hat{J}_y \hat{U}^{\dagger}_c(\chi).
                \end{align}
         Note that the choice of operators $\hat{J}_y$ and $\hat{J}_z$ in \eqref{eq:cubic_unitary} and \eqref{eq:operator_O} is not unique - they could be swapped with no bearing on generality. To complete the definition of squeezing, both specific and resource, we need to define the free states and the free operations. In harmonic oscillators, the goal of cubic operation is to go beyond the Gaussian domain and cubic squeezing can be taken as a witness of non-Gaussianity. It is not as straightforward in the systems of collective spins, but we can separate the target state from a class of experimentally feasible states defined as ground states of Hamiltonian. We define Gaussian-like states as ground states of the Hamiltonian
                \begin{align}\label{eq:Hamiltonian}
                    \begin{split}
                        \hat{H}(g)  = \hat{U}_F&\left[  \left( \frac{g^2}{1+g^2}\right) \hat{J}^2_z \right. +\left(\frac{1}{1+g^2}\right)\hat{J}^2_y\\&+ \left. \left(\frac{g}{1+g^2}\right)\hat{J}_x \right]\hat{U}_F^{\dagger}.
                    \end{split}
                \end{align}
        This Hamiltonian, based on the general twist and turn Hamiltonian \cite{Muessel2015Aug}, represents two types of operations depending on the choice of the $g$ parameter: two-axis counter-twisting (TACT) operation for $g\in (0,\infty)\setminus\{1\}$ and one-axis twisting (OAT) operation for  $g \in \{0,1\}$. The free operations $\hat{U}_F$ represent a rotation of the Bloch sphere along an arbitrary axis - an operation that cannot create any non-trivial property of the state.

        The ground states of \eqref{eq:Hamiltonian} cover the entire class of squeezed Gaussian-like states, including the coherent state -- the linear term in the Hamiltonian \eqref{eq:Hamiltonian} takes off the degeneracy, so the eigenstates are only on one hemisphere of the Bloch sphere. Gaussian-like states are states in the Holstein-Primakoff approximation \cite{Holstein1940Dec} that belong to the class of Gaussian states in harmonic oscillator phase space. From now on, we will use the term Gaussian and non-Gaussian states to mean Gaussian-like and non-Gaussian-like states.

        We can now discuss some specific applications of the defined cubic operator, in the increasing order of numerical cost.
            \begin{itemize}
            \item[(i)]
            \textbf{Direct evaluation of specific squeezing.} For any given state, the squeezing in a given cubic nonlinear operator can be evaluated by calculating the variance of (\ref{eq:operator_O}). This is checked against the minimal value obtainable by the free states to see whether this state is relevant for potential applications. 
                
                \begin{equation}\label{eq:nonlinear_var_basic}
                    \xi_{\chi} = \frac{\text{min}_{\hat{U}_F}\left[\text{var}_{\hat{\rho}}\left( \hat{U}_F\hat{\mathcal{O}}_{c_{\chi}}\hat{U}^{\dagger}_F \right)\right]}{\text{min}_{\hat{U}_F}\text{min}_{\hat{\rho}_{F}}\left[\text{var}_{\hat{\rho}_{F}}\left(\hat{U}_F\hat{\mathcal{O}}_{c_{\chi}}\hat{U}^{\dagger}_F\right)\right] }.
                \end{equation}
            Note that in some cases it is also possible to look at the eigenstates of the operator restricted to some subspace to quickly get the best representation of the ideal state \cite{Konno2021Feb, Fiurasek2022Aug, Marek2024May}.

                \item[(ii)]
            \textbf{Deeper characterization of \textit{resource} squeezing.} In experiments, it is often important to witness whether some broader quantum feature is present in the state reconstructed from the experimental data in case there was some previously unaccounted influence. In this case, we could be interested in cubic squeezing with any parameter $\chi$, prompting a modification of \eqref{eq:resource variance}. To search for which cubic parameter is the variance of the tested state minimal, we minimize the entire fraction over the possible nonlinear interaction strength $\chi$. The resulting relation has the form
                \begin{equation}\label{eq:nonlinear_var_minchi}
                    \xi = \text{min}_{\chi}\frac{\text{min}_{\hat{U}_F}\left[\text{var}_{\hat{\rho}}\left( \hat{U}_F\hat{\mathcal{O}}_c(\chi)\hat{U}^{\dagger}_F \right)\right]}{\text{min}_{\hat{U}_F}\text{min}_{\hat{\rho}_{F}}\left[\text{var}_{\hat{\rho}_{F}}\left(\hat{U}_F\hat{\mathcal{O}}_c(\chi)\hat{U}^{\dagger}_F\right)\right] }.
                \end{equation}

            \item[(iii)]
            \textbf{State preparation.} Finally, we can be interested in optimizing the state preparation circuit with the goal of obtaining the quantum state with the desired property. Rather than searching in the full set of quantum states, we might be searching in a subset of states given by specific preparation procedure \cite{Brauer2021Jul}. In this case, we need to minimize the cubic variance for any given state from the set and then minimize over the states from the set:
                \begin{equation}\label{eq:nonlinear_var_minstate}
                    \xi_{\hat{\mathcal{O}}_{\chi}} = \frac{\text{min}_{\hat{\rho}}\text{min}_{\hat{U}_F}\left[\text{var}_{\hat{\rho}}\left( \hat{U}_F\hat{\mathcal{O}}_{\chi}\hat{U}^{\dagger}_F \right)\right]}{\text{min}_{\hat{U}_F}\text{min}_{\hat{\rho}_{F}}\left[\text{var}_{\hat{\rho}_{F}}\left(\hat{U}_F\hat{\mathcal{O}}_{\chi}\hat{U}^{\dagger}_F\right)\right] }.
                \end{equation}
            \end{itemize}

            In case (ii) and (iii), the formula contains unitary operations $\hat{U}_F$ through which we minimize. We always solve this step by centering the tested state in one preselected axis (in the given axis we rotate the state so that its variance matrix is diagonal), in which we calculate the given variance. Details of this calibration are contained in the Supplement material.

            \subsection{Example of a superposition of two Dicke states}

                As an illustrative example, we chose to test a state that is defined as a superposition of two Dicke states with a free parameter $\gamma$
                    \begin{equation}\label{eq:0+1}
                        \ket{\psi} = \sqrt{\gamma}\ket{0} + \sqrt{(1 - \gamma)}\ket{1},
                    \end{equation}
                where the Dicke states are defined with the help of collective spin states as: $\ket{0} = \ket{\frac{N}{2},-\frac{N}{2}}$  and $\ket{0} = \ket{\frac{N}{2},-\frac{N}{2} + 1}$. This state has been used as an approximation of a cubically squeezed state \cite{Konno2021Feb, Miyata2016Feb, Kala2022Aug}. The reason is simple; this state is consistent both with different numbers of atoms in the system and in a harmonic oscillator -- this is demonstrated in the supplement.

\begin{figure}
                        \centering
                        \includegraphics[width=8.5cm]{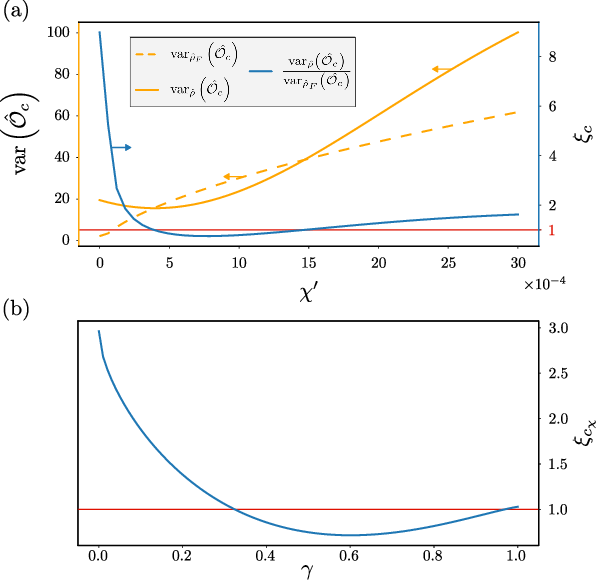}
                        \caption{Numerical evaluation of cubic squeezing for superposition of the first two Dicke states \eqref{eq:0+1} in systems with $N=80$. (a) The variance of the cubic nonlinear operator \eqref{eq:nonlinear_var_minchi} relative to the effective cubicity $\chi'$. The yellow dashed curve corresponds to the variance of the benchmark state; the yellow full line corresponds to the variance of the analyzed state $\eqref{eq:0+1}$ with $\gamma = \frac{1}{2}$ (the yellow curves are connected to the left $y$-axis, which is also yellow). The blue curve describes the resulting parameter of equation \eqref{eq:nonlinear_var_minchi} (the values of the blue curve correspond to the right $y$-axis, which is also blue), the red curve is the limit corresponding to $\xi_c = 1$, i.e. when we are below this limit, the state is non-linearly cubically squeezed when above the limit, we achieve better values with the benchmark states. (b) Optimal nonlinear squeezing  \eqref{eq:nonlinear_var_minstate} relative to the parameter gamma of the state $\gamma$ for the state described by equation \eqref{eq:0+1}. The blue curve is the optimal nonlinear squeezing parameter $\xi$ obtainable for the particular state. For $\gamma = \frac{1}{2}$ the value of $\xi$ corresponds to the minimum of the blue curve in Fig.\ref{fig:img1}(a). The red curve denotes $\xi=1$ that delineates the boundary between Gaussian and non-Gaussian.}
                        \label{fig:img1}
                    \end{figure}

                  \textit{ Deeper characterization of the resource squeezing} - we chose a particular state \eqref{eq:0+1}, one with $\gamma = \frac{1}{2}$ and $N = 80$, and found its optimal cubic squeezing by minimizing the squeezing parameter \eqref{eq:nonlinear_var_minchi} over the possible nonlinear interaction strengths $\chi$. As can be seen in Fig. \ref{fig:img1}(a), the state has $\xi<1$ and is therefore cubically squeezed and non-Gaussian for $4\times10^{-4}<~\chi'<~15\times10^{-4} $ $\left( \chi' = \frac{\chi}{3}\left[\frac{N}{2}\right]^{-\frac{3}{2}} \right)$. The maximal cubic squeezing corresponds to the minimum $\xi = 0.748$ ($-1.26$ dB), and it is found for $\chi' = 7.96\times10^{-4}$.

                 \textit{State preparation} - for this task we try to optimize the resource squeezing for a state specified by some constraints on preparation. This can be a set of defined operations \cite{Brauer2021Jul}, but in this instance, we will simply consider an arbitrary state with the form \eqref{eq:0+1}. The goal now is to find $\gamma$, for which the resource squeezing has minimal value. The results are shown in Fig. \ref{fig:img1}(b). The value for $\gamma = \frac{1}{2}$ corresponds to the minimum in Fig. \ref{fig:img1}(a) and we can immediately see that it is not the full optimum. The optimum is $\gamma = 0.551$, and we achieve squeezing $\xi_c = 0.715$, which corresponds to $-1.459$dB. $\chi'$ parameter was set $\chi' = 7.959 \times 10^{ -4}$ to match the previously used optimal value.
                    
        \section{Conclusion}
           We present a novel perspective on state evaluation, which focuses on the use of variance-based squeezing to observe specific state properties. By selecting an appropriate operator for variance analysis, we can reveal distinct characteristics of quantum states. In addition, for any class of benchmarking states we chose, the variance is bounded. This means that we can use this variance to verify whether the examined state can be a part of this class or whether it has to be excluded. This is of particular interest when the states from the benchmarking class are unsuitable for some task, as is the case for, for example, separable, classical, or Gaussian states.

            Our approach is first discussed in general terms and then demonstrated with an example involving a discrete system of collective spins. Specifically, we explore cubic compression in the superposition of Dicke states $\ket{0}$ and $\ket{1}$. The discussion highlights various ways in which variance can be utilized, both for evaluating quantum states and for their preparation. We believe that this result will find its application in the evaluation of experimental data and finding optimal states for use in quantum metrology and quantum information processing.

        \section{Acknowledgements} We acknowledge Grant No. 22-08772S of the Czech Science Foundation, the European Union’s HORIZON Research and Innovation Actions under Grant Agreement No. 101080173 (CLUSTEC), and funding from the Ministry of Education, Youth and Sport of the Czech Republic (CZ.02.01.01/00/22 008/0004649). SB acknowledges project IGA PrF-2024-008.

\newpage
\clearpage

\begin{onecolumngrid}

\noindent
    {\centering
    {\bfseries \large Supplement material for: Generalized squeezing as a witness\\[1em]}
    \normalsize \v{S}imon Br\"auer, Tom\'a\v{s} Opatrn\'y, Petr Marek\\
    \it Department of Optics, Faculty of Science, Palack\'{y} University,
    \\ 17. Listopadu 12, 77146 Olomouc, Czech Republic\\[1em]
    }

In the following we will show that the cubic nonlinear squeezing, which manifests as reduced variance of operator
    \begin{align}\label{eq:operator_O_supp}
	\hat{\mathcal{O}}_c(\chi) = \hat{U}_c(\chi) \hat{J}_y \hat{U}^{\dagger}_c(\chi),
\end{align}
where
   \begin{align}\label{eq:cubic_unitary_supp}
	\hat{U}_c (\chi) = \exp{\left(i\frac{\chi}{3} \left(\frac{N}{2}\right)^{-\frac{3}{2}} \hat{J}_z^3\right)},
\end{align}
is consistently described across the different sizes of the investigated collective spin systems. The size of a system is given by the number of atomic spins, $N$, and the relevant range, in which we can try to observe the nonlinear squeezing lies between $N=4$ and $N\rightarrow \infty$ when the systems converges into the Holstein-Primakoff approximation of continuous variables linear harmonic oscillator (LHO). It is not meaningful to evaluate cubic squeezing for $N\le 3$, when investigating the non-Gaussian like properties given by parameter
    \begin{equation}\label{eq:nonlinear_var_minchi_supp}
	\xi = \text{min}_{\chi}\frac{\text{min}_{\hat{U}_F}\left[\text{var}_{\hat{\rho}}\left( \hat{U}_F\hat{\mathcal{O}}_c(\chi)\hat{U}^{\dagger}_F \right)\right]}{\text{min}_{\hat{U}_F}\text{min}_{\hat{\rho}_{F}}\left[\text{var}_{\hat{\rho}_{F}}\left(\hat{U}_F\hat{\mathcal{O}}_c(\chi)\hat{U}^{\dagger}_F\right)\right] },
\end{equation}
because the Gausssian-like operations are expressed as eigenstates of Hamiltonian quadratic in the momentum operators. For  $N\le 3$, the space is spanned by only four Dicke states $\ket{0},\ket{1},\ket{2},\ket{3}$, and the third power of angular momentum operators is not meaningfully different from the second power. 

For the illustration, we have evaluated the nonlinear squeezing (\ref{eq:nonlinear_var_minchi_supp}) for a test state given by superposition of the first two Dicke states,
    \begin{equation}\label{eq:0+1_supp}
                        \ket{\psi} = \sqrt{\frac{1}{2}}\ket{0} + \sqrt{\frac{1}{2}}\ket{1},
    \end{equation}
which is an approximation of the cubically squeezed state. The results are shown in figure \ref{fig:img2}. We can see that the value of nonlinear squeezing parameter $\xi$ is lowest for the smallest system with $N=4$ and that it steadily grows until it reaches the limit for the Harmonic oscillator. The largest changes happen in the interval $N \in [4,100]$, which is marked by the gray area in the figure. Beyond this point, the value of squeezing differs by only 0.015 from the LHO case (which corresponds to 1.95$\%$ difference). 		

The reason why the nonlinear squeezing of the test state is so good for the small sizes of the system is that in those dimensions, the first two Dicke states of the test state span $40\%$ of the total Hilbert space of the system, and the test state can therefore best match the optimal state, which is the minimal value eigenstate of \eqref{eq:operator_O_supp}. As the dimension of the system increases, this influence vanishes. This behavior is illustrated in figure \ref{fig:spheres_with_groundstates}, where we can see differences between the test state and the ideal state for different dimensions of the system.  

    \begin{figure}[ht!]
            \centering
            \includegraphics[width=15cm]{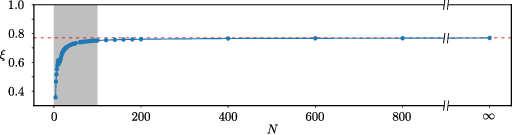}
            \caption[]{Results of numerical simulation of minimizing nonlinear squeezing parameter with test state \eqref{eq:0+1_supp} in the interval of $N$, where the grey area corresponds to the $N\in[4,100]$,  $\infty$ point in the $x$-axis corresponds to the transition to the harmonic oscillator phase space.}
            \label{fig:img2}
    \end{figure}   
    
    \begin{figure}[ht!]
    	\centering
    	\includegraphics[width=0.8\textwidth]{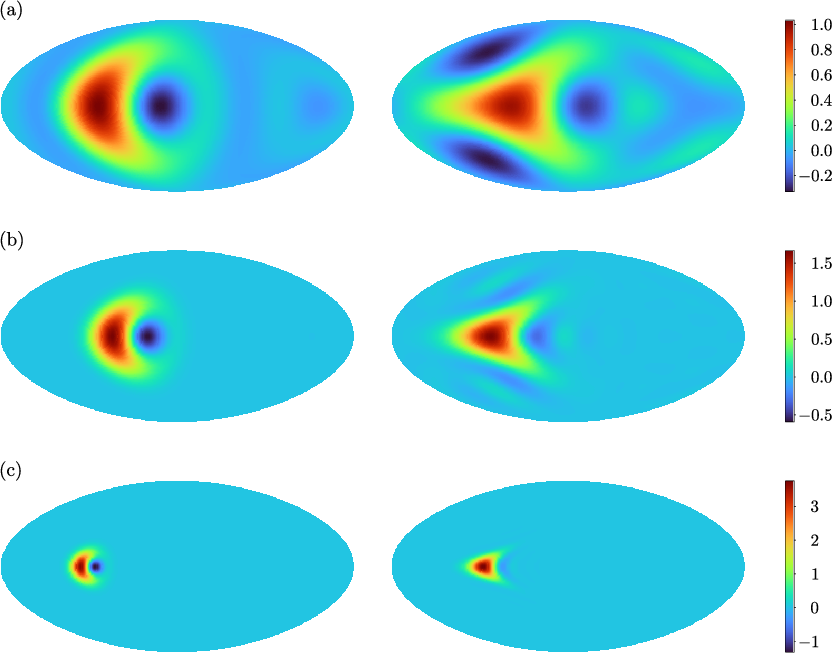}
    	\caption[Hammers projection of the Bloch spheres with Wigner function of the states of collective spins.]{ Bloch spheres in Hammers projection. The left column corresponds to the tested state \eqref{eq:0+1_supp}, i.e. the superposition of $\ket{0}$ and $\ket{1}$ for different numbers of atoms in the systems (where $\gamma = 1/2$). The right column shows the ground states of the operator $\hat{\mathcal{O}}$ \eqref{eq:operator_O_supp} for systems with different numbers of atoms. (a) $N = 4$ (b) $N = 10$ (c) $N = 50.$}
    	\label{fig:spheres_with_groundstates}
    \end{figure}
\end{onecolumngrid}
\end{document}